\documentclass[prl,twocolumn,showpacs,preprintnumbers,amsmath,amssymb]{revtex4}
\usepackage{graphicx}% Include figure files

\usepackage{dcolumn}% Align table columns on decimal point
\usepackage{bm}% bold math

\begin{document}

\newcommand*{\cm}{cm$^{-1}$\,}
\newcommand*{\Tc}{T$_c$\,}
\newcommand*{\EuFeAs}{EuFe$_2$As$_2$\,}

%\reprint{APS/123-QED}

\title{Measurement of the c-axis optical reflectance of AFe$_2$As$_2$ (A=Ba, Sr)
single crystals: Evidence of different mechanisms for the
formation of two energy gaps}% Force line breaks with \\

\author{Z. G. Chen}
\author{T. Dong}
\author{R. H. Ruan}
\author{B. F. Hu}
\author{B. Cheng}
\author{W. Z. Hu}
\author{P. Zheng}
\author{Z. Fang}
\author{X. Dai}
\author{N. L. Wang}

\affiliation{Beijing National Laboratory for Condensed Matter
Physics, Institute of Physics, Chinese Academy of Sciences,
Beijing 100190, China}

\begin{abstract}
We present the c-axis optical reflectance measurement on single
crystals of BaFe$_2$As$_2$ and SrFe$_2$As$_2$, the parent
compounds of FeAs based superconductors. Different from the
ab-plane optical response where two distinct energy gaps were
observed in the SDW state, only the smaller energy gap could be
seen clearly for \textbf{E}$\parallel$c-axis. The very pronounced
energy gap structure seen at a higher energy scale for
\textbf{E}$\parallel$ab-plane is almost invisible. We propose a
novel picture for the band structure evolution across the SDW
transition and suggest different driving mechanisms for the
formation of the two energy gaps.

\end{abstract}

\pacs{74.25.Gz, 74.70.Xa, 75.30.Fv}


\maketitle

For quasi-two dimensional layered materials, striking differences
could exist in the in-plane and out-of-plane charge transport and
dynamics. For example, in some high-T$_c$ cuprates, metallic
in-plane charge transport coexists with nonmetallic conductivity
along the c-axis\cite{Ando,Uchida}. The contrasting behavior
violates the conventional concept of band electron transport, and
has been the subject of intensive study. Fe-pnictide
superconducting materials also crystalize in the layered structure
with Fe-As layers separated by alkaline metal ions or other
insulator-like layers. Band structure calculations based on the
local-density approximation (LDA) or generalized gradient
approximations (GGA) indicate dominantly two-dimensional (2D)
cylinder-like Fermi surfaces (FSs) along the
c-axis\cite{Lebegue,Singh,Ma}. It is important to see whether or
not the Fe-pnictides share similar anisotropic charge dynamical
properties with cuprates.

Optical spectroscopy is a powerful technique to investigate charge
dynamics and band structure of a material as it probes both free
carriers and interband excitations. In particular, it yields
direct information about the energy gap formation in the broken
symmetry state. Optical spectroscopy studies on the ab-plane
properties of different Fe-pnictides and chalcogenides systems
have been reported by several
groups.\cite{Li,Hu122,Pfuner,Yang,DWu,Qazilbash,GFChen1,Hu111,Akrap,Moon,Chen1111,Heumen,Kim}
For the parent compounds of Fe-pnictides, the measurements provide
clear evidence for the formation of the partial energy gaps in the
magnetic phase, supporting the itinerant picture that the energy
gain for the antiferromagnetic ground state is achieved through
the opening of a spin-density-wave (SDW) gap on the
FSs.\cite{Hu122,DWu,Akrap,Moon,Chen1111} For the superconducting
samples, the superconducting pairing gaps were also detected by
the technique.\cite{Li,Heumen,Kim} However, optical investigations
have not been carefully done on the c-axis response of Fe-pnictide
materials. There is only one work in the literature containing
optical data along the c-axis.\cite{DWu} Unfortunately, the data
were limited to the high frequencies, above 700 \cm. Because of
this limitation, neither the free-carrier response nor any feature
related to the SDW gap were observed. In fact, the reported
reflectance data appear to have extraordinarily low values. As
information about the anisotropic charge dynamics is extremely
important for understanding the materials, a detailed and careful
determination of the c-axis optical response is highly necessary.

In this letter we present the c-axis
(\textbf{E}$\parallel$\textbf{c}) optical reflectance measurement
over broad frequencies on thick BaFe$_2$As$_2$ and SrFe$_2$As$_2$
single crystal samples. We observed the SDW energy gap formation
in the low-T magnetic ordered state. However, different from the
ab-plane response where two distinct energy gaps were identified
for AFe$_2$As$_2$ (A=Ba, Sr), only the gap corresponding to the
smaller energy scale of \textbf{E}$\parallel$ab-plane could be
clearly seen for the polarization parallel to the c-axis. The more
pronounced gap structure at the higher energy scale for
\textbf{E}$\parallel$ab-plane becomes almost invisible. The
significant difference between the two polarizations has important
implication for the electronic structure of those compounds. A
schematic picture for the band structure evolution across the
structural/magnetic transition was proposed to understand the
experimental findings.

Large-sized single crystals of BaFe$_2$As$_2$ and SrFe$_2$As$_2$
were grown from the FeAs flux in Al$_2$O$_3$ crucibles sealed in
quartz tubes. The growth procedure is similar to the description
in our earlier work \cite{Chen2} except for one major difference:
the crucible and quartz tube were placed in a direction of 45
degree relative to the vertical. After completing the growth
procedure and breaking the crucible, we can easily find relatively
thick crystals growing from the inner surface of the crucible
along the direction at 45 degree relative to the crucible cylinder
axis (i.e. crystals grow vertically). The dc resistivity measured
by the four contact technique on cleaved ab-plane was found to be
almost identical to the data presented in our early work, showing
a sharp drops at 138 K and 200 K for BaFe$_2$As$_2$ and
SrFe$_2$As$_2$, respectively, being ascribed to the formation of
SDW order.\cite{Hu122} We performed x-ray diffraction measurement
to check the single crystalline nature of our samples and their
orientation. The crystals were then cut in a direction
perpendicular to the cleaved ab-plane. The cutting surfaces were
finely polished for the c-axis polarization measurement.

The optical reflectance measurements with
\textbf{E}$\parallel$c-axis were performed on a Bruker IFS 66v/s
spectrometer in the frequency range from 50 to 25000 cm$^{-1}$. An
\textit{in situ} gold and aluminium overcoating technique was used
to get the reflectivity R($\omega$). The real part of conductivity
$\sigma_1(\omega)$ is obtained by the Kramers-Kronig
transformation of R($\omega$). A Hagen-Rubens relation was used
for low frequency extrapolation. A $\omega^{-0.5}$ dependence was
used for high frequency extrapolation up to 300000 \cm, above
which a $\omega^{-4}$ dependence is employed.

\begin{figure}
\includegraphics[width=7.5cm,clip]{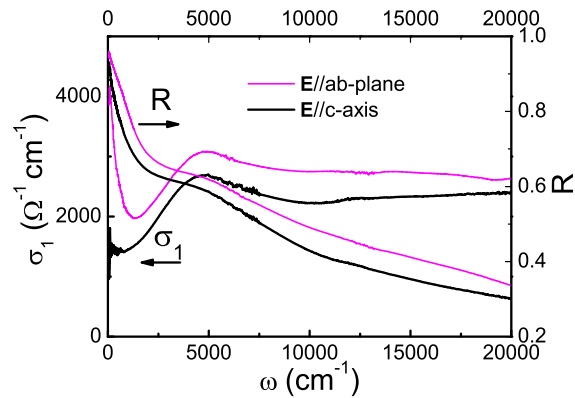}
%\centerline{\includegraphics[width=3.2in]{fig1.eps}}%
\caption{\label{fig:EuLaR} The c-axis R($\omega$) and
$\sigma_1(\omega)$ at 300 K for BaFe$_2$As$_2$ over a broad
frequencis up to 20000 \cm. The ab-plane spectra were also
included for a purpose of comparison.\cite{Hu122} The conductivity
anisotropy at low frequency limit is about 2.8.}
\end{figure}

Figure 1 shows the c-axis R($\omega$) and $\sigma_1(\omega)$ at
room temperature over a broad frequency range up to 20000 \cm for
BaFe$_2$As$_2$. For a comparison, we also plot the optical spectra
with \textbf{E}$\parallel$ab-plane\cite{Hu122}. We can see that
the overall R($\omega$) along the c-axis is quite similar to that
in the ab-plane except for relatively lower values. This is
dramatically different from the optical spectra of some high-Tc
cuprates\cite{Uchida} and other layered compounds, for example,
layered ruthenates\cite{Katsufuji}, where the c-axis R($\omega$)
shows much lower values and quite different frequency-dependent
behavior from the ab-plane. This observation suggests that the
band structure of Fe-pnictide should be quite three-dimensional
(3D), in contrast to the expectation based on its layered crystal
structure. The difference in $\sigma_1(\omega)$ spectra between
ab-plane and the c-axis seems to become larger at low frequencies.
An extrapolation to zero frequency, or the dc conductivity, shows
an anisotropy ratio of about 2.8. This matches well with the
anisotropy ratio determined directly from the dc resistivity
measurement by a careful Montgomery technique (about
3$\pm$1)\cite{Tanatar}. We noticed that the c-axis reflectance
values obtained here are much higher than the data presented by Wu
et al. on a EuFe$_2$As$_2$ sample\cite{DWu}. The reflectance
values in their measured frequency range (above 700 \cm) are
already below 0.4, leading to extremely low values of conductivity
(below 100 $\Omega^{-1}$cm$^{-1}$ at 700 \cm). We remark here that
we repeated measurement of \textbf{E}$\parallel$\textbf{c} on
another crystal grown from a different batch and achieved almost
identical result.

\begin{figure}[t]
\begin{center}
\includegraphics[width=6.5cm,clip]{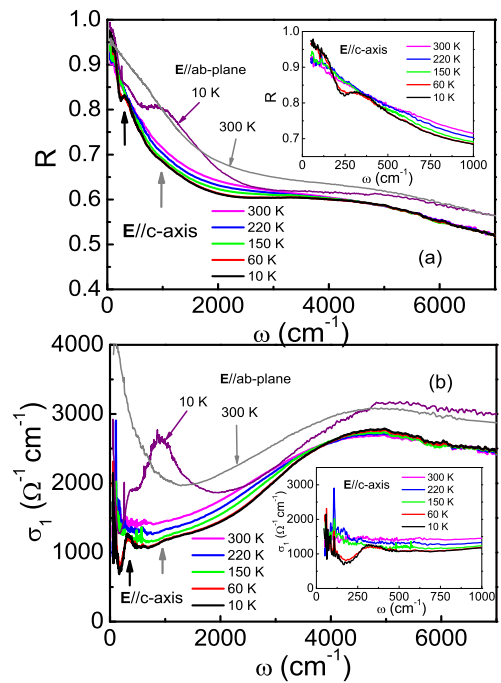}
%\includegraphics[width=3.2in]{fig2.eps}%
%\vspace*{-0.20cm}%
\caption{\label{fig:EuRandSandRho} The c-axis R($\omega$) (a) and
$\sigma_1(\omega)$ (b) at different temperatures below 7000 \cm
for BaFe$_2$As$_2$. Inset: expanded plot of the low-$\omega$
R($\omega$) and $\sigma_1(\omega)$ spectra. The ab-plane
R($\omega$) and $\sigma_1(\omega)$ at 300 K and 10 K are included
for comparison. The two upward arrows indicate the anomalies
associated with the SDW order. The very pronounced peak in
$\sigma_1(\omega)$ for \textbf{E}$\parallel$ab-plane becomes
extremely weak for \textbf{E}$\parallel$\textbf{c}.}
\end{center}
\end{figure}

The temperature dependences of the R($\omega$) and
$\sigma_1(\omega)$ spectra of BaFe$_2$As$_2$ below 7000 \cm
($\sim$1eV) are plotted in Fig. 2 (a) and (b), respectively. In
the mid-infrared region (near 4000 \cm), there is a strong
suppression feature in both R($\omega$) and $\sigma_1(\omega)$
spectra. Essentially, the same feature is seen in the ab-plane
optical response at slightly higher frequencies (near 5000 \cm).
For a comparison we also plot the in-plane R($\omega$) and
$\sigma_1(\omega)$ spectra at 300 K and 10 K in the
figure.\cite{Hu122} The suppressed spectral weight in
$\sigma_1(\omega)$ is largely transferred to the high frequencies
above 5000 \cm for both polarizations. The suppression is a common
feature for Fe-pnictide materials and is not directly related to
the SDW order as it is seen well above the structural/magnetic
transition temperature.\cite{Hu122,Hu111,Chen1111} Its origin
remains to be explored. Most remarkably, the reflectance is
strongly suppressed below $\sim$320 \cm upon entering the SDW
ordered state. A sharp upturn appears at lower energy scale. The
strong suppression in R($\omega$) results in a prominent gap
structure in $\sigma_1(\omega)$. The sharp upturn at lower
frequencies in R($\omega$) leads to a narrow residual Drude
component at very low frequencies (see the inset of Fig. 2 (a) and
(b)). Above T$_{SDW}$, the Drude feature is not clear for
\textbf{E}$\parallel$c-axis, being different from the spectra in
the ab-plane. The data provide clear evidence for the partial
energy gap formation along the c-axis in the SDW ordered phase as
well. In the ab-plane optical spectra,\cite{Hu122} we observed two
energy gaps in the SDW state with the conductivity peak energies
near 360 \cm and 890 \cm. Here we found that the energy scale of
the gap for \textbf{E}$\parallel$c is close to the smaller one
seen in the ab-plane $\sigma_1(\omega)$. The more pronounced gap
structure at higher energy in the ab-plane becomes extremely weak
for the c-axis polarization in both R($\omega$) and
$\sigma_1(\omega)$, as indicated by an upward grey arrow in the
figure.

\begin{figure}[t]
\begin{center}
\includegraphics[width=7.0cm,clip]{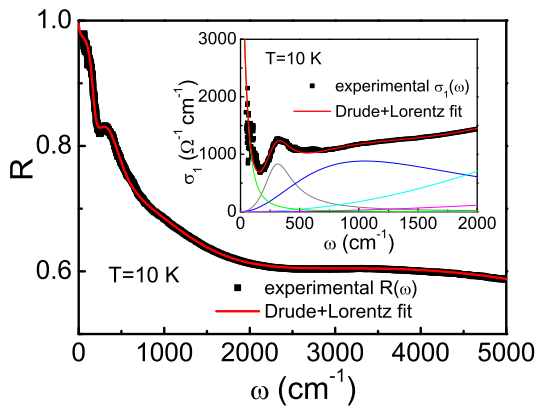}

%\includegraphics[width=3.2in]{fig3.eps}%
%\vspace*{-0.20cm}%
\caption{\label{fig:EuRandSandRho} The experimental R($\omega$)
(main panel) and $\sigma_1(\omega)$ (inset) at 10 K of
BaFe$_2$As$_2$ for \textbf{E}$\parallel$\textbf{c} and the fitting
curves from a Drude-Lorentz model up to different energy scales at
10 K. The same parameters were used to reproduce both R($\omega$)
and $\sigma_1(\omega)$. The decomposed Drude component and the
Lorentz peaks are also shown in the inset.}
\end{center}
\end{figure}

It is tempting to analyze the data in a more quantitative way.
Since a residual narrow Drude component is present at very low
temperature in the SDW state, it is relatively easy to isolate its
spectral weight. We use a standard Drude-Lorentz model to
decompose the spectra into different components:
\begin{equation}
\epsilon(\omega)=\epsilon_\infty-{{\omega_p^2}\over{\omega^2+i\omega/\tau}}+\sum_{i=1}^N{{S_i^2}\over{\omega_i^2-\omega^2-i\omega/\tau_i}}.
\label{chik}
\end{equation}
Here, $\epsilon_\infty$ is the dielectric constant at high energy,
the middle and last terms are the Drude and Lorentz components,
respectively. The experimental R($\omega$) and $\sigma_1(\omega)$
spectra at 10 K could be well reproduced by using a Drude and
several several Lorentz peaks, as shown in Fig. 3\cite{Remark}.
This analysis leads to $\omega_{pc}$=3600 \cm and 1/$\tau_c$=47
\cm for the Drude component at 10 K. Our earlier study on the
ab-plane properties indicates that those two parameters are 4660
\cm and 55 \cm at 10 K, respectively.\cite{Hu122} Then we find
that the anisotropy ratio of the spectral weight, which is equal
to the ratio of the square of plasma frequency
$\omega_{pc}^2$/$\omega_{pab}^2$, is about 60$\%$ at 10 K. The
value indicates that the anisotropy of the band structure
reconstructed in the SDW order state is not very strong. Since the
Drude-like features above T$_{SDW}$ are not very clear, there is
no unique or unambiguous way to isolate the Drude components from
the overall spectra, we did not try to estimate those parameters
at high temperatures.

\begin{figure}[t]
\begin{center}
\includegraphics[width=6.5cm,clip]{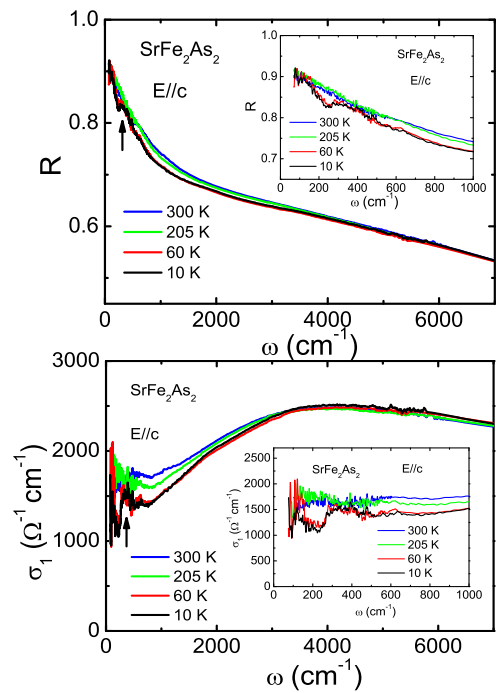}
%\includegraphics[width=3.2in]{fig2.eps}%
%\vspace*{-0.20cm}%
\caption{\label{fig:EuRandSandRho} The c-axis R($\omega$) and
$\sigma_1(\omega)$ at different temperatures below 7000 \cm for
SrFe$_2$As$_2$. Inset: expanded plot of the low-$\omega$
R($\omega$) and $\sigma_1(\omega)$ spectra. The upward arrows
indicate the anomalies associated with the SDW order.}
\end{center}
\end{figure}

The substantial difference of low frequency gap structures in the
SDW state between the two polarizations has important implication
for the electronic structure of Fe-pnictides. It is important to
check whether the observations are generic for 122-type parent
compounds. For this purpose, we performed optical measurement on
SrFe$_2$As$_2$ single crystals. Because the obtained
SrFe$_2$As$_2$ crystals have smaller dimension along the c-axis,
the signal-to-noise ratio at low frequencies is much smaller.
Nevertheless, we observed essentially the same spectral features
for SrFe$_2$As$_2$ as well. As shown in Fig. 4, only the smaller
gap could be seen for \textbf{E}$\parallel$c-axis in the SDW
state, the large energy gap is almost invisible. Those experiments
indicate that the observed spectral features are generic for
122-type undoped compounds.

\begin{figure}[t]
\begin{center}
\includegraphics[width=8.5cm,clip]{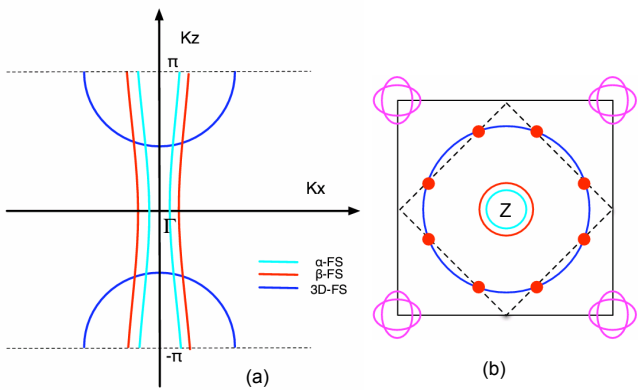}
%\includegraphics[width=3.2in]{fig4.eps}%
%\vspace*{-0.20cm}%
\caption{\label{fig:EuRandSandRho} (a) The schematic plot of the
Fermi surfaces along the k$_z$ direction around the $\Gamma$
point. Besides the 2D FSs ($\alpha$ and $\beta$ bands) there is an
additional 3D FS enclosing the Z point ((0,0,$\pi$)). (b) A top
view of the FSs. The magnetic Brilliouin zone driven by the
($\pi,\pi$) nesting between the disconnected 2D cylinder-like FSs
would cut the 3D FS and open gaps in the intersecting points.}
\end{center}
\end{figure}

Understanding why two distinct gaps could be observed for
\textbf{E}$\parallel$ab-plane but only a smaller one could be
clearly seen for \textbf{E}$\parallel$c-axis is a crucial issue
here. It has important implication for the band structure
evolution across the transition. We found that our data could be
naturally explained by assuming that there are two types of FSs in
the system: the 2D cylinder like FSs and a large-size 3D ellipsoid
like FS, as schematically presented in Fig. 5. Presence of the
large-size 3D ellipsoid FS with dominant Fe-3d$_{3z^2-r^2}$
orbital is indicated by the recent {\it ab-initio} LDA+Gutzwiller
calculations, where electron correlations are taken into account
beyond LDA \cite{Wang}. Recent transport\cite{Tanatar}, upper
critical field\cite{Yuan}, as well as ARPES
experiments\cite{Liu,Malaeb} also suggest the presence of a 3D FS.

If we neglect the small dispersion of those 2D FSs along the
c-axis, the Fermi velocity along the c-axis becomes zero and the
electrons on the 2D FSs would only couple to the polarized light
with \textbf{E}$\parallel$ab-plane. On the other hand, the
electrons on the 3D FS can couple to the light polarized along any
direction. We suggest that the 2D cylinder-like hole FSs around
$\Gamma$ and the 2D electron FSs around the corner of the
Brillouin zone are strongly nested, which is the main driving
force for the SDW instability of the system, while the 3D
ellipsoid FS does not show any nesting with other FSs.
Nevertheless, with the formation of the SDW order driven by the
($\pi,\pi$) nesting of 2D FSs, the new magnetic Brillouin zone
boundary would cut the large-size 3D ellipsoid FS and result in an
energy gap at the intersecting points as shown in Fig. 5 (b). This
can be seen by optics in both \textbf{E}$\parallel$ab-plane and
\textbf{E}$\parallel$\textbf{c}. Then we ascribe the larger gap
seen most clearly for \textbf{E}$\parallel$ab-plane to the
nesting-driven gap opened on 2D FSs, the smaller one to the gap
formed on 3D FS which could be considered as the consequence of
the SDW order. Because the 2D FSs are more dramatically affected
by the SDW instability than the 3D FS, we expect a reduction of
the anisotropy in the magnetic ordered state.

To summarize, we have successfully grown thick single crystals of
AFe$_2$As$_2$ (A=Ba, Sr) and investigated their optical
properties. Our study revealed a clear difference in optical
conductivity for \textbf{E}$\parallel$ab-plane and
\textbf{E}$\parallel$c-axis in the SDW state. The very pronounced
energy gap structure seen at a higher energy scale for
\textbf{E}$\parallel$ab-plane is almost invisible for
\textbf{E}$\parallel$c-axis, whereas the smaller energy gap could
be seen in both polarizations. We propose a novel picture for the
band structure evolution and suggest different driving mechanisms
for the two energy gaps. The strong ($\pi,\pi$) nesting between
disconnected 2D cylinder-like FSs is the main driving force for
the SDW instability, leading to the opening of larger energy gap
in the 2D FSs. The cutting of the magnetic Brillouin zone on the
3D FS leads to a smaller gap at the crossing region.

\begin{acknowledgments}
This work is supported by the National Science Foundation of
China, the Knowledge Innovation Project of the Chinese Academy of
Sciences, and the 973 project of the Ministry of Science and
Technology of China.

\end{acknowledgments}


\begin{thebibliography}{99}
\bibitem{Remark} The low- and mid-infrared R($\omega$) and $\sigma_1(\omega)$ could be
well reproduced by one Drude and three Lorentz components centered
at 320, 1000, and 4500 \cm (for SDW gap and interband
transitions). To reproduce the spectra upto 20000 \cm, one extra
Lorentz term at higher energy has to be added.
\end{thebibliography}
\end{document}